# 3D printing of hierarchical structures made of inorganic silicon-rich glass featuring self-forming nanogratings

Po-Han Huang[1], Shiqian Chen[2], Oliver Hartwig[3], David E. Marschner[1], Georg S. Duesberg[3], Göran Stemme[1], Jiantong Li[2], Kristinn B. Gylfason[1], and Frank Niklaus[1,*]

[1] Division of Micro and Nanosystems, School of Electrical Engineering and Computer Science, KTH Royal Institute of Technology, Stockholm 10044, Sweden

[2] Division of Electronics and Embedded Systems, School of Electrical Engineering and Computer Science, KTH Royal Institute of Technology, Kista 16440, Sweden

[3] Institute of Physics, EIT 2, Faculty of Electrical Engineering and Information Technology, University of the Bundeswehr Munich & SENS Research Center, Neubiberg 85577, Germany

* Corresponding author: Frank Niklaus (frank@kth.se)

## Abstract

Hierarchical structures are abundant in nature, such as in the superhydrophobic surfaces of lotus leaves and the structural coloration of butterfly wings. They consist of ordered features across multiple size scales, and their unique properties have attracted enormous interest in wide-ranging fields, including energy storage, nanofluidics, and nanophotonics. Femtosecond lasers, capable of inducing various material modifications, have shown promise for manufacturing tailored hierarchical structures. However, existing methods such as multiphoton lithography and 3D printing using nanoparticle-filled inks typically involve polymers and suffer from high process complexity. Here, we demonstrate 3D printing of hierarchical structures in inorganic silicon-rich glass featuring self-forming nanogratings. This approach takes advantage of our finding that femtosecond laser pulses can induce simultaneous multiphoton crosslinking and self-formation of nanogratings in hydrogen silsesquioxane (HSQ). The 3D printing process combines the 3D patterning capability of multiphoton lithography and the efficient generation of periodic structures by the self-formation of nanogratings. We 3D-printed micro-supercapacitors with large surface areas and a remarkable areal capacitance of 1 $mF/cm^2$ at an ultrahigh scan rate of 50 V/s, thereby demonstrating the utility of our 3D printing approach for device applications in emerging fields such as energy storage.

## Introduction

Hierarchical structures that possess ordered features with dimensions at the micro- and nanoscales offer unique properties such as large surface area[1], structural coloration[2–6], custom surface wetting[3,5,7], negative mechanical Poisson's ratio[8], and defined porosity[1,4,9–13]. Thus, hierarchical structures are highly relevant for applications in energy storage[1,9–11] (e.g., supercapacitors and batteries), catalysis[14], photonics[2–6,12,13,15], photovoltaic devices[16], fluidics[3,7,13], mechanical metamaterials[8], and biology[5]. Fabrication of micro- and nanoscale hierarchical structures using femtosecond laser direct writing has attracted enormous interest because it offers unprecedented resolution and design flexibility[2–8,13–15,17]. At the focal point of a femtosecond laser beam, the photon density can be sufficiently high to cause multiphoton absorption in the exposed material volume. Multiphoton absorption can, in turn, induce localized material modifications, including (1) material crosslinking[18] and (2) formation of self-organized structures[4,19]. On the one hand, (1) material crosslinking by multiphoton absorption has been applied to additive manufacturing of 3D structures by direct writing of photoresists with nanoscale resolution, known as multiphoton lithography[18]. Multiphoton lithography has been used extensively for 3D printing of hierarchical structures but requires direct writing of every single nanoscale feature[8,20], which requires long patterning times and complex laser-writing paths to optimize the exposure of each nanoscale feature. Alternatively, photoresists for multiphoton lithography have been pre-loaded with nanoscale particles that self-assemble during printing to fabricate 3D hierarchical structures[2]. However, incorporating nanoscale particles within photoresists is typically challenging since many parameters, such as rheology, printability, material and process compatibility, and homogeneity, critically affect the process. In addition, the nanoscale structures obtained with this approach have the same shape throughout the 3D-printed object, and their orientation and distribution cannot be accurately controlled across the different parts of the object. Furthermore, the materials amenable to multiphoton lithography are largely limited to organic polymers that cannot resist high temperatures and have limited chemical stability. Recently, multiphoton lithography has been applied to 3D printing of inorganic materials[21–26]. However, these approaches still require long patterning times and complex laser-writing paths to fabricate 3D hierarchical structures. On the other hand, (2) multiphoton absorption in the focal point of a femtosecond laser beam at higher photon fluxes can induce the formation of self-organized structures. This effect has been observed inside glasses[27–32] and on the surfaces of dielectrics, metals, and semiconductors[17,19]. The effect has been applied to fabricating hierarchical structures with tailored optical, mechanical, and fluidic properties[3–5,13] that are relevant for important application fields such as structural coloration[5], nanophotonics[4], data storage[15], tribology[33], and fluidics[5]. However, due to the subtractive nature of the formation processes of these self-organized structures, their application has been constrained to material surfaces or structures embedded inside transparent materials, resulting in limited design freedom and integration flexibility. The possibility to induce simultaneous (1) multiphoton crosslinking and (2) formation of self-organized structures in a material upon femtosecond-laser exposure could facilitate efficient fabrication of 3D hierarchical structures, which has not been demonstrated to date.

Here, we report an approach for 3D printing of hierarchical structures in inorganic silicon-rich (Si-rich) glass. This is achieved by additive manufacturing of femtosecond-laser-induced self-organized nanogratings, taking advantage of our finding that a single femtosecond-laser exposure inside hydrogen silsesquioxane (HSQ) can simultaneously cause both (1) crosslinking of HSQ into Si-rich glass and (2) formation of self-organized nanogratings in the printed Si-rich glass. Our material characterization of the resulting nanogratings shows that the constituent nanoplates feature a core of silica-like glass surrounded by silicon clusters and silicon-rich glass species. Moreover, our experiments indicate that the printed nanogratings share the formation mechanism with the nanogratings observed inside bulk glass materials upon femtosecond laser exposure. Uniquely, the morphology of our printed nanogratings can be controlled by the polarization of the femtosecond laser beam and by the input laser dose. With these capabilities, our approach enables the rapid fabrication of micro-supercapacitors (MSCs) with abundant open channels in the electrodes that provide a large surface area and facilitate fast ion transport. As a demonstration, we 3D-printed integrated MSCs with remarkable performance at ultrahigh scan rates that are promising for emerging applications, such as energy storage in self-powered electronics.

## Results

### 3D printing of hierarchical Si-rich glass structures

Our approach for 3D printing of hierarchical Si-rich glass structures consists of three main steps: (1) drop-casting of HSQ on a substrate, (2) 3D printing by femtosecond laser direct writing in the HSQ, and (3) development of the laser-written patterns in aqueous potassium hydroxide (KOH) solution (Fig. 1a, b). The 3D structures printed with this approach feature self-forming periodic nanogratings (Fig. 1). Specifically, a single line written with the femtosecond laser in the HSQ and subsequently developed in KOH, results in a line structure composed of multiple overlapping nanogratings, with each nanograting consisting of multiple nanoplates that are equally spaced, aligned to each other, and partially connected (Fig. 1a-c and Supplementary Fig. 1). There are three controllable levels within the hierarchy of the 3D-printed structures: (i) the 3D architecture, (ii) the self-organized nanogratings, and (iii) the nanoplates (Fig. 1b, c). The overall shape of the printed 3D architecture (level (i)) can be freely defined by the laser writing paths. As for the nanoplates (level (iii)), their orientation is controllable by the polarization of the femtosecond laser, i.e., the orientation of the electric field of the laser light, while the thickness of the nanoplates is fixed. The nanoplates are always perpendicular to the laser polarization, regardless of the direction of the laser writing paths (Fig.

1c, d). Therefore, the orientation of the nanoplates in every part of the 3D architecture can be chosen freely and independently by simply tuning the laser polarization during laser writing (Fig. 1d-f). We observed that the nanoplates were 0.86 ± 0.11 μm thick without observable dependence on the laser polarization or the exposure dose. The exposure dose is the total input laser energy in a unit length of laser writing paths and depends on the laser pulse energy and the number of input laser pulses during printing. As for the nanogratings (level (ii)), their dimensions and alignments are controllable, while the grating periodicity is fixed, and the orientation of the subordinate nanoplates determines the grating orientation. For a nanograting resulting from the laser writing path of a single line, the total width of the nanograting increased with the exposure dose (Fig. 1g). The increase in nanograting width resulted from an increase in the number of subordinate nanoplates. At the same time, the width of every nanoplate and the nanograting periodicity remained at 0.86 ± 0.11 μm and 1.06 ± 0.08 μm, respectively. The smallest width of a single-line nanograting that we realized was approximately 800 nm, consisting of only one nanoplate that was written using a laser power slightly above the power threshold for printing, with the laser polarization being perpendicular to the direction of the laser writing path that formed the line (Fig. 1g and Supplementary Fig. 2). Moreover, we demonstrated the possibility of extending nanogratings in 3D by stitching multiple laser writing paths with the same laser polarization which resulted in the self-alignment of neighboring nanogratings (Fig. 1e, f). This capability enables printing large 3D architectures with controlled orientation and uniform periodicity of the subordinate nanogratings throughout the architectures.

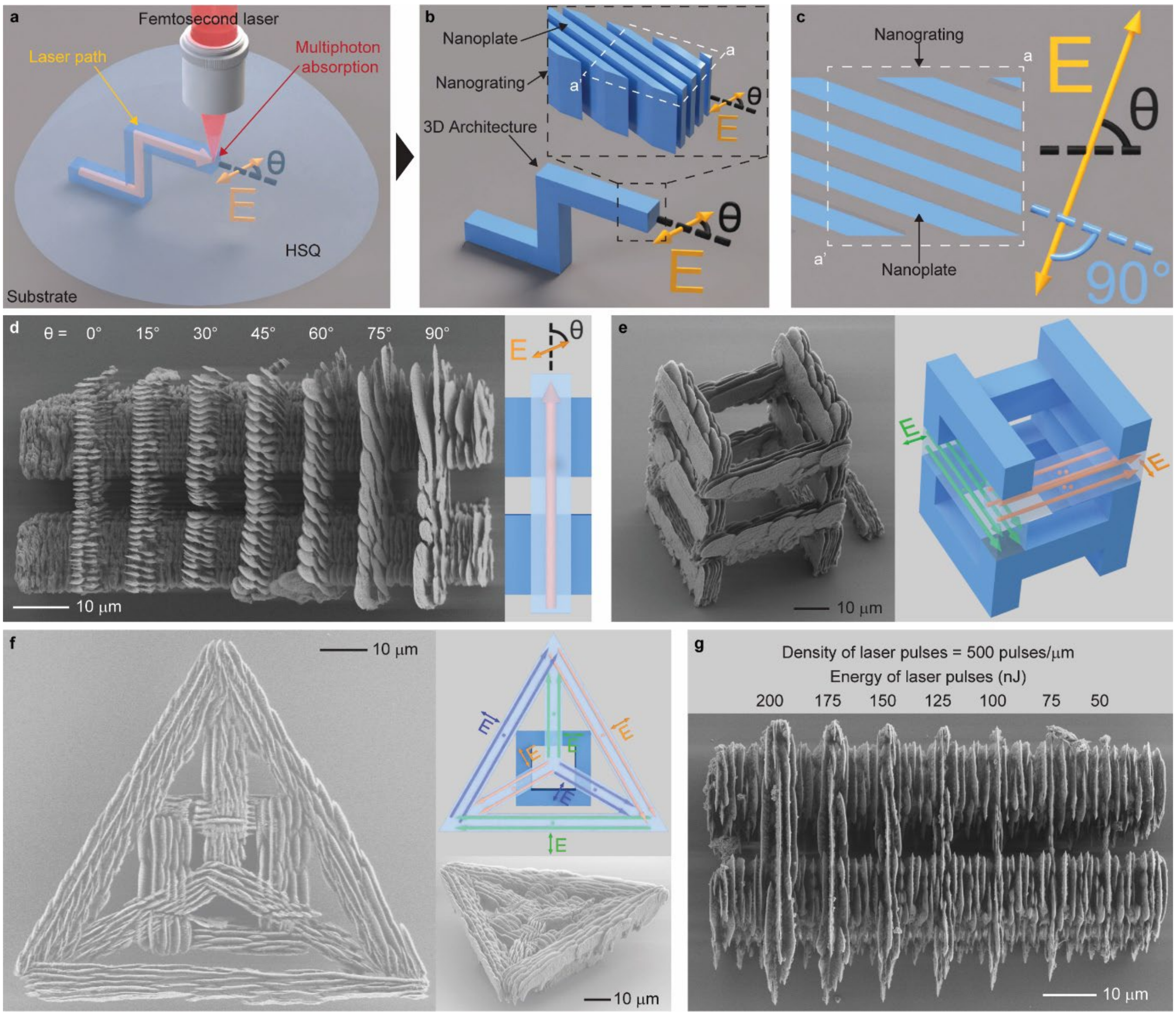

**Fig. 1. 3D printing of self-organized hierarchical structures in inorganic Si-rich glass.** **a,** Schematic illustration of femtosecond laser direct writing inside drop-casted HSQ. The arrow inside the HSQ represents the laser writing path. The chosen orientation of the electric field E, i.e., the laser polarization, is indicated with a short double-sided arrow of matching color. **b,** Illustration of the self-organized hierarchy ((i) 3D architecture, (ii) nanograting, and (iii) nanoplates) inside the 3D-printed glass architectures using the laser fabrication scheme shown in **a**, and after development in a KOH solution. **c,** Enlarged top view of the structure in **b** illustrating that the orientation of the nanoplates is perpendicular to the laser polarization. The nanoplates are partially connected with the neighboring nanoplates, which for simplicity is not depicted in the illustration. **d,** SEM image and laser fabrication scheme of 3D-printed suspended single-line structures printed with different angular offsets θ between the polarization and the writing direction. **e, f,** SEM images and laser fabrication schemes of a 3D-printed scaffold (**e**) and a suspended triangle (**f**), demonstrating the capability of locally tuning the orientation of the nanogratings and extending the nanogratings over large areas. **g,** SEM image of 3D-printed suspended single-line structures printed using laser pulses with different energies of between 50 and 200 nJ with the same spatial input laser pulse density of 500

pulses/μm.

**Laser-pulse-number-dependent nanograting formation**

To elucidate the formation mechanism of the self-organized nanogratings in the 3D-printed structures, we fabricated single-spot structures with different numbers of femtosecond laser pulses. During the fabrication of each single-spot structure, the laser focal point was static at one position inside the HSQ. In this way, the effects of the femtosecond laser pulses on the exposed HSQ can be observed independently of the effects of the dynamic laser writing process. We printed two sets of single-spot structures with orthogonal laser polarizations. Each set included four single-spot structures exposed by 1, 5, 20, and 100 laser pulses, respectively, and the temporal separation of the laser pulses was fixed to one millisecond (Fig. 2a). The SEM images of the resulting single-spot structures after development showed that their morphology depends on the number of input laser pulses and that the laser polarization only affects the orientation of the morphology. With only one pulse, the structures showed no nanoplate feature. As the number of input laser pulses increased to five, nanoplates that were oriented perpendicularly to the laser polarization appeared in the single-spot structures. However, the dimensions of the nanoplates and the distances between them within the individual single-spot structures varied. When the number of input laser pulses increased to 20, the shape of the nanoplates became sharper, and their dimensions and periodicity became more homogeneous. Further increasing the number of input laser pulses up to 100 resulted in slightly more homogeneous and sharper nanoplates. These observations show that the formation of the nanogratings in the 3D-printed glass structures is a self-forming and gradual process involving tens of laser pulses. Moreover, the formation process is not an instant reaction that requires laser energy to be deposited within a specific short time span since the temporal separation of each input laser pulse we used was one millisecond. One millisecond is much longer than the time scale of the laser pulse duration (femtoseconds) or the expected lifetime of the excitons resulting from laser-glass interactions (microseconds)[34]. Self-formation of polarization-dependent nanogratings has been previously observed inside different bulk glass materials after femtosecond laser exposure, and the formation processes were also found to be gradual and dependent on the number of input laser pulses[28–30,35,36]. Therefore, the nanograting formation mechanism in HSQ should be comparable to the one observed in bulk glass materials. However, the formation process of nanogratings inside transparent materials upon femtosecond laser exposure has not been completely understood, although several physical models involving light, plasma, or excitons have been proposed[4,30,37,38]. In our 3D printing process, the pulse-number-dependent nanograting formation offers the possibility to selectively control the sharpness and regularity of the nanoscale features inside the different parts of the 3D-printed architectures. The higher the density of input laser pulses in a 3D-printed volume, the sharper the nanoplates and the better the alignment of the nanoplates within the volume (Fig. 2b-e).

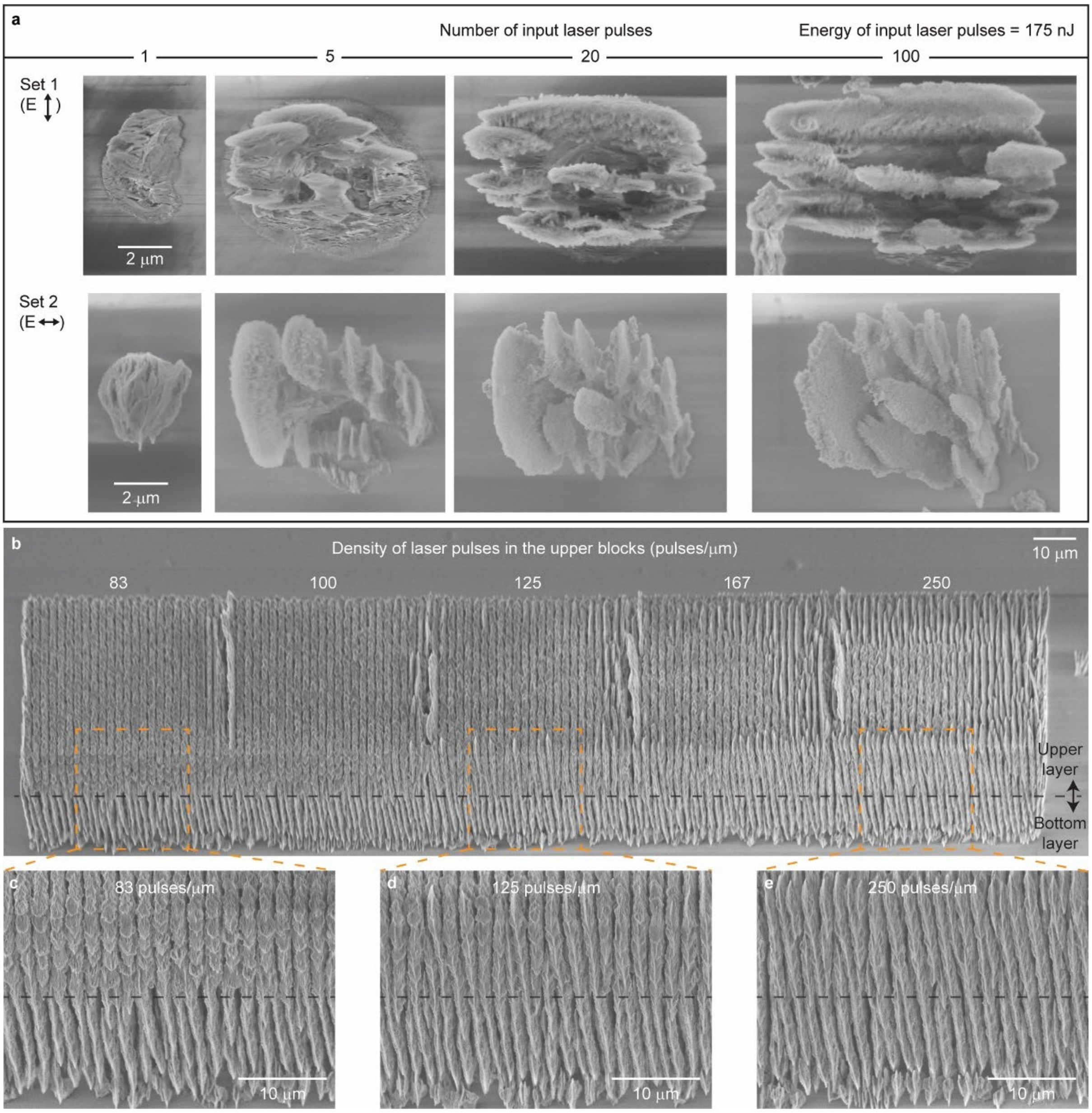

**Fig. 2. Dependence of the structural appearance of the printed nanogratings on the number of input laser pulses. a,** SEM top-view images of two sets of single-spot structures fabricated with orthogonal laser polarizations (marked as E in the image). Each single-spot structure in a set was printed with a different number of laser pulses, while the laser focus was static at one spot in the material. The energy and temporal separations of each laser pulse were fixed at 175 nJ and 1 ms, respectively. Both sets of single-spot structures include four single-spot structures exposed to 1, 5, 20, and 100 laser pulses, respectively. The images in each set are scaled relative to the scale bar in the 1-pulse structure image within that set. **b,** Tilted-view SEM image of a 3D-printed two-layer structure. The bottom layer was printed with a spatial density of input laser pulses of 500 pulses/μm, while each block in the upper layer was printed using different densities of input laser pulses as annotated. The spatial densities of input laser pulses refer to the pulse densities in each laser-written line, while the entire structure was

printed by overlapping multiple lines with a fixed line density. The laser pulse repetition rate was fixed at 50 kHz, and the laser scanning speed was varied to achieve different laser pulse densities. The interface between the two layers is marked with a dashed line. The subordinate nanoscale structures in the blocks are less regular and sharp with a lower spatial input laser pulse density. **C, d, e,** Enlarged tilted-view SEM images of the side walls of the corresponding blocks in **b**. The interface between the upper layer and bottom layer in each image is marked with a dashed line.

**Underlying mechanisms of the 3D printing process**

To elucidate the underlying mechanisms of our proposed 3D printing process, we characterized the elemental composition and chemical bonds of the 3D-printed glass by energy-dispersive X-ray spectroscopy (EDS), Raman spectroscopy, and photoluminescence (PL) spectroscopy (Fig. 3 and Supplementary Section 1). By analyzing the EDS, Raman, and PL spectra (Fig. 3), and our morphological observations in the 3D printed structures, we infer that the printed material is Si-rich oxide glass with residual hydrogen ($H_nSiO_x$, $n < 1$ and $x < 1.5$) transformed from HSQ by the femtosecond-laser exposure. The transformation of HSQ to Si-rich oxide glass involves two distinct mechanisms: (1) in-volume nanograting formation in bulk glass materials[27–32,36], and (2) cage-to-network transformation in HSQ[24,39]. Mechanism (1) refers to the formation of polarization-dependent self-organized periodic patterns inside the femtosecond-laser-exposed volume of bulk glass materials, including silica glass[27,36], porous glass[30], multicomponent silicate glasses[28,29], sapphire[32], and germania glass[31]. Such patterns have been previously observed in a format of periodically distributed oxygen deficiencies[27,31,36] or periodically distributed cracks[30]. Mechanism (2) refers to the crosslinking of HSQ resulting from femtosecond-laser exposure during which the chemical structure of HSQ transforms from the pristine cage form with a low chemical etch resistance, to the crosslinked network form with a high chemical etch resistance, as has been reported recently[24,39]. Regarding mechanism (1), first, the periodic morphology of our printed glass and its dependence on the laser polarization agree well with observations of in-volume nanogratings inside bulk glass materials (Fig. 1)[27–32,36]. Moreover, an increased content of 3- and 4-membered ring structures in the silica network and the appearance of non-bridging oxygen hole centers (NBOHC) seen in the Raman and PL spectra, respectively, have also been commonly observed in femtosecond-laser-exposed silica glass[40] (Fig. 3c, d). The non-bridging oxygen hole centers have been proposed as the basis of nanograting formation[35]. However, our observation that the femtosecond-laser-exposed HSQ survived the KOH development (i.e., etching), which resulted in the 3D-printed glass structure, is inconsistent with mechanism (1). This is because femtosecond-laser-induced nanogratings in bulk glass materials involve only material phases that have decreased or unchanged chemical etch resistance to hydrofluoric acid (HF)[28,29,32,36] and KOH[41] as compared to pristine glass materials. Our observation of an increased chemical etch resistance of the femtosecond-laser-exposed HSQ indicates the occurrence of the mechanism (2) cage-to-network transformation (crosslinking) in HSQ. After being transformed from the cage form to the network form, HSQ is known to have an increased chemical etch resistance and a lowered intensity of the characteristic Raman peak of Si-H at ~2260 $cm^{-1}$ due to the associated hydrogen dissipation[42]. The latter is well supported by the absence of the Si-H peak in the spectrum of our printed glass (Supplementary Fig. 4). However, the crosslinked HSQ should have a higher oxygen content than the pristine cage HSQ[24,43], which does not agree with the decrease of the O/Si ratio in the femtosecond-laser-exposed HSQ seen in our EDS data (Fig. 3a). To further elucidate our observations and the considerations above, we characterized the femtosecond-laser-exposed regions when they were still embedded in pristine HSQ using SEM and EDS (Fig. 3e-g). To do so, we patterned a region inside HSQ and mechanically cleaved the sample along the plane perpendicular to the orientation of the formed nanoplates in the patterned region. The cross-section of the sample was characterized directly after cleaving without polishing or chemical etching. In the SEM images of the cross-section, we observed periodically distributed cracks in the femtosecond-laser-exposed regions (Fig. 3e). Moreover, the EDS elemental maps of those regions showed that the O/Si ratio of the area inside the cracks is significantly lower than in the materials surrounding the cracks (Fig. 3f, g). The EDS signals observed in the areas inside the cracks likely contained contributions from the surfaces surrounding the cracks, considering a slight tilt of the cross-section relative to the EDS beam direction, and considering the width of the cracks being not perfectly even.

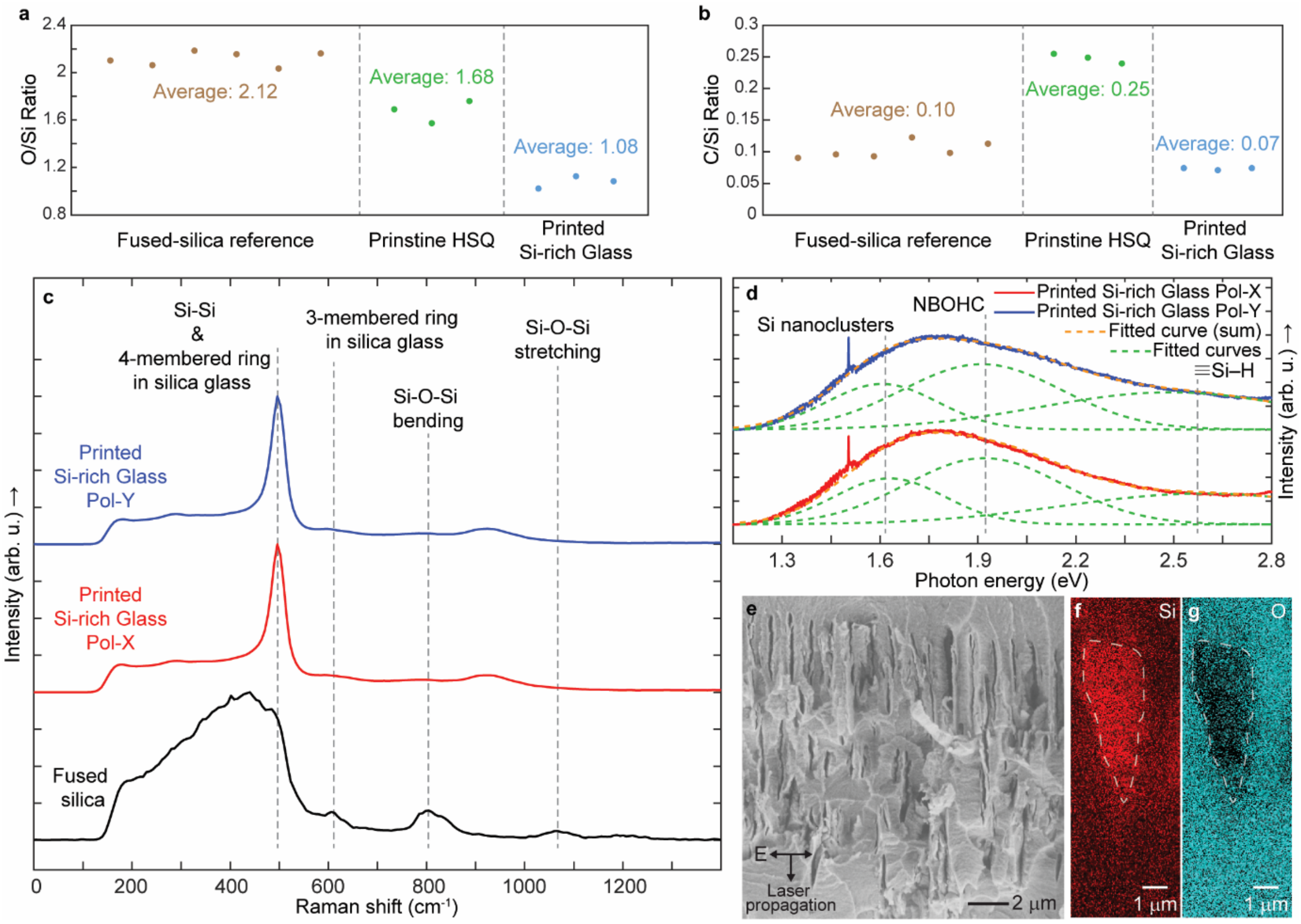

**Fig. 3. Characterization of the 3D-printed Si-rich glass. a, b,** Atomic ratio of oxygen to silicon (O/Si) and carbon to silicon (C/Si), respectively, measured in the pristine HSQ, the 3D-printed Si-rich glass (i.e., femtosecond-laser-exposed HSQ), and a fused silica substrate as a reference using energy-dispersive X-ray spectroscopy. Each data point was integrated from an area of at least 1300 $\mu m^2$. The data shows that the oxygen and carbon contents in HSQ were decreased during its transformation to Si-rich glass. **c,** Raman spectra measured in the fused silica substrate (black) and in two Si-rich glass samples printed with the laser polarization in parallel (Pol-X, red) and perpendicular (Pol-Y, blue) to the laser writing direction. The Si-rich glass spectra of the two samples show no difference and indicate that the laser polarization does not affect the material properties of the printed Si-rich glass. **d,** Photoluminescence spectra measured in the two Si-rich glass samples in **c**, which also show no difference. **e,** SEM image

of the cross-section of a mechanically cleaved sample containing 3D-printed Si-rich glass embedded inside HSQ. No polishing or chemical etching was used. Periodically distributed cracks can be observed. The materials between the cracks form the nanogratings after development. **f, g,** Representative EDS elemental maps in atomic percent of silicon (Si) and oxygen (O), respectively, measured in the region surrounding a crack in the 3D-printed Si-rich glass shown in **e**. Regions with higher atomic percentages of silicon or oxygen appear brighter. The boundaries of the crack are marked with dashed lines. The O/Si ratio is lower inside the crack, which was observed in all the cracks we have characterized (more than 20 cracks).

Taken together, these observations are consistent with the following underlying mechanisms of our 3D printing process: At the start, the pristine HSQ is mainly in the cage form. Upon exposure to one femtosecond-laser pulse, the center part of the exposed region of HSQ gains sufficient energy through multiphoton absorption to crosslink and transform from the pristine cage form to the crosslinked network form (mechanism (2), Fig. 2a). Consequently, that part of the laser-exposed HSQ has an increased chemical etch-resistance to KOH. During the same exposure, the energy at some locations within the exposed region exceeds the threshold for the decomposition of Si-O bonds, which results in the formation of nanovoids associated with the release of $O_2$ gas, which has been observed in silica glass (mechanism (1))[44]. The following laser pulses further crosslink the entire exposed region of HSQ (mechanism (2)), which is supported by our observation of the increased size of single-spot structures with an increasing number of input laser pulses (Fig. 2a). At the same time, those laser pulses extend the nanovoids into periodically distributed nanocracks (mechanism (1))[30,44]. The Si-O bonds at the location of the nanocracks were decomposed into $O_2$ gas and Si-rich species, which can include Si-rich oxides and Si clusters. This corresponds well with the EDS elemental maps that show a reduced O/Si ratio inside the nanocracks (Fig. 3f, g) and the observation of Raman and PL peaks that are related to Si-Si bonds and Si nanoclusters (Fig. 3c, d). This can also explain the low overall O/Si ratio measured in the 3D-printed glass (Fig. 3a), which is essentially composed of nanoplates made of crosslinked HSQ ($H_nSiO_x$, $n < 1$ and $x > 1.5$) with their surfaces covered by Si-rich species.

**3D-printed micro-supercapacitors**

To demonstrate the utility of our approach for 3D printing of hierarchical structures in inorganic Si-rich glass, we printed and evaluated on-chip micro-supercapacitors (MSCs). On-chip MSCs are promising energy-storage devices that can be combined with energy-harvesting components to realize autonomous and maintenance-free microelectronic systems, due to the long lifetimes and high power densities of MSCs[45]. Our 3D printing approach is well suited for the rapid fabrication of high-performance on-chip integrated MSCs for two reasons. First, our approach enables precise control over the morphology of the printed architectures across multiple size scales and hierarchical levels. This capability is beneficial for fabricating 3D MSC architectures with a large surface area that results in a large total capacitance, abundant internal channels designed to facilitate fast ion transport, and small footprints[9,10]. Second, our printed Si-rich glass features exceptional chemical and thermal stability and long-term durability. These properties are crucial for functionalizing the printed architectures with desired electrochemically active materials and enabling the MSCs to operate continuously and in harsh environments. We realized two different MSCs for our demonstration, each with a small footprint of 280 µm by 200 µm, by 3D printing hierarchical structures as the skeletons of the MSC electrodes on silicon substrates (Fig. 4a, b). The high-level 3D architecture of both MSCs was identical. Each consisted of two electrodes, and each electrode consisted of 33 vertical Si-rich glass sheets mechanically supported by a horizontal connecting bar printed on top. However, the nanoscale structure of the two MSCs was different. In the first MSC, the nanogratings in the vertical sheets were designed to have short nanoplates oriented perpendicular to the extended direction of the sheets (Fig. 4a, named ⊥-MSC hereafter). In the second MSC, the nanoplates were made long and oriented in parallel to the extended direction of the vertical sheets to which they belonged (Fig. 4b, named ǁ-MSC hereafter). After printing, the MSC skeletons were coated with a 25 nm thick layer of titanium nitride (TiN) using atomic-layer deposition (ALD). This layer served as the electrode material and current collector.

To characterize the electrochemical performance of the MSCs, we drop-casted a gel electrolyte consisting of a poly(4-styrenesulfonic acid)/LiCl composite onto the MSCs and performed cyclic voltammetry (CV) at scan rates between 5 V/s and 100 V/s. In the CV curves, both MSCs showed considerably larger areal capacitances than the reference, which is a TiN-coated flat substrate without a 3D-printed structure (Fig. 4c), confirming that our 3D-printed hierarchical architectures provided a considerably increased total surface area. Moreover, the CV curves of both MSCs remained close to rectangular at scan rates as high as 50 V/s, indicating their high-rate capabilities. However, the dependences of the performance of the ⊥-MSC and ||-MSC on the scan rate were significantly different. At scan rates below 15 V/s, the MSCs showed comparable areal capacitances, which indicates that their effective total surface areas were comparable (Fig. 4d). On the other hand, at scan rates above 15 V/s, the areal capacitance of the ||-MSC decreased dramatically with an increasing scan rate. In contrast, the areal capacitance of the ⊥-MSC remained relatively stable at about 1.0 $mF/cm^2$. This observation implies that the ion transport was faster in the ⊥-MSC than in the ||-MSC. To validate this, we plot the logarithm of the charging current $i$ at the voltage of 0.5 V in the CV curves against the logarithm of the scan rate $v$ (Fig. 4e). This plot can be used to study the behaviors of MSCs using the power law $i \sim v^b$ where the value of the exponent $b$ provides insight into the charge-storage kinetics[9,10,46]. A $b$ value close to 1 indicates that the kinetics of the MSC are dominated by the high-rate capacitive storage mechanism, while a $b$ value close to 0.5 indicates that the kinetics of the MSC are dominated by the slow diffusion-limited storage mechanism[9]. By fitting our plots with the power law model, we obtained the $b$ values for the ⊥-MSC and the ||-MSC which, are 0.96 and 0.73, respectively, in the range of scan rates between 15 and 100 V/s. These results confirm that the ion transport was much faster in the ⊥-MSC than in the ||-MSC

(Fig. 4e). The difference in the ion transport rate clearly demonstrated the strong influence of the nanoscale structure of the MSCs on their performance. The nanoplates in the ⊥-MSC were short and discrete, which left abundant and large open channels for fast ion transport, while the nanoplates in the ∥-MSC were long and relatively enclosed, resulting in lengthy and tortuous channels that hinder ion transport (Fig. 4f)[9,10,46]. Moreover, in contrast to the randomly oriented ion channels of most devices reported in literature[47], the open channels in our ⊥-MSC were vertically aligned from the top to the bottom of the electrodes thanks to the excellent controllability of our 3D printing process. Such unique "through" channels enable our ⊥-MSC to attain a large areal capacitance of 1.0 mF/cm$^2$ at a high scan rate of 50 V/s even with the solid-state gel electrolytes that have lower ionic conductivity than liquid electrolytes. Remarkably, our ⊥-MSC has outperformed most high-rate MSCs reported in the literature (Supplementary Table 1). Furthermore, to demonstrate the stability of our Si-rich glass MSCs in high-temperature and long-cycling conditions, we performed CV characterization with an ionic-liquid electrolyte of 1-butyl-3-methylimidazolium tetrafluoroborate at a scan rate of 5 V/s up to a temperature of 200 °C and 10000 charging-discharging cycles. The CV curves measured at room temperature remained rectangular with a retention of 83% of capacitance up to 10000 cycles, showing its long-cycling stability (Supplementary Fig. 5a, b). Furthermore, our MSCs remained functional up to 200 °C with an areal capacitance of 2.6 mF/cm$^2$, demonstrating these MSCs' potential for emerging electronics applications in harsh environments (Supplementary Fig. 5c, d).

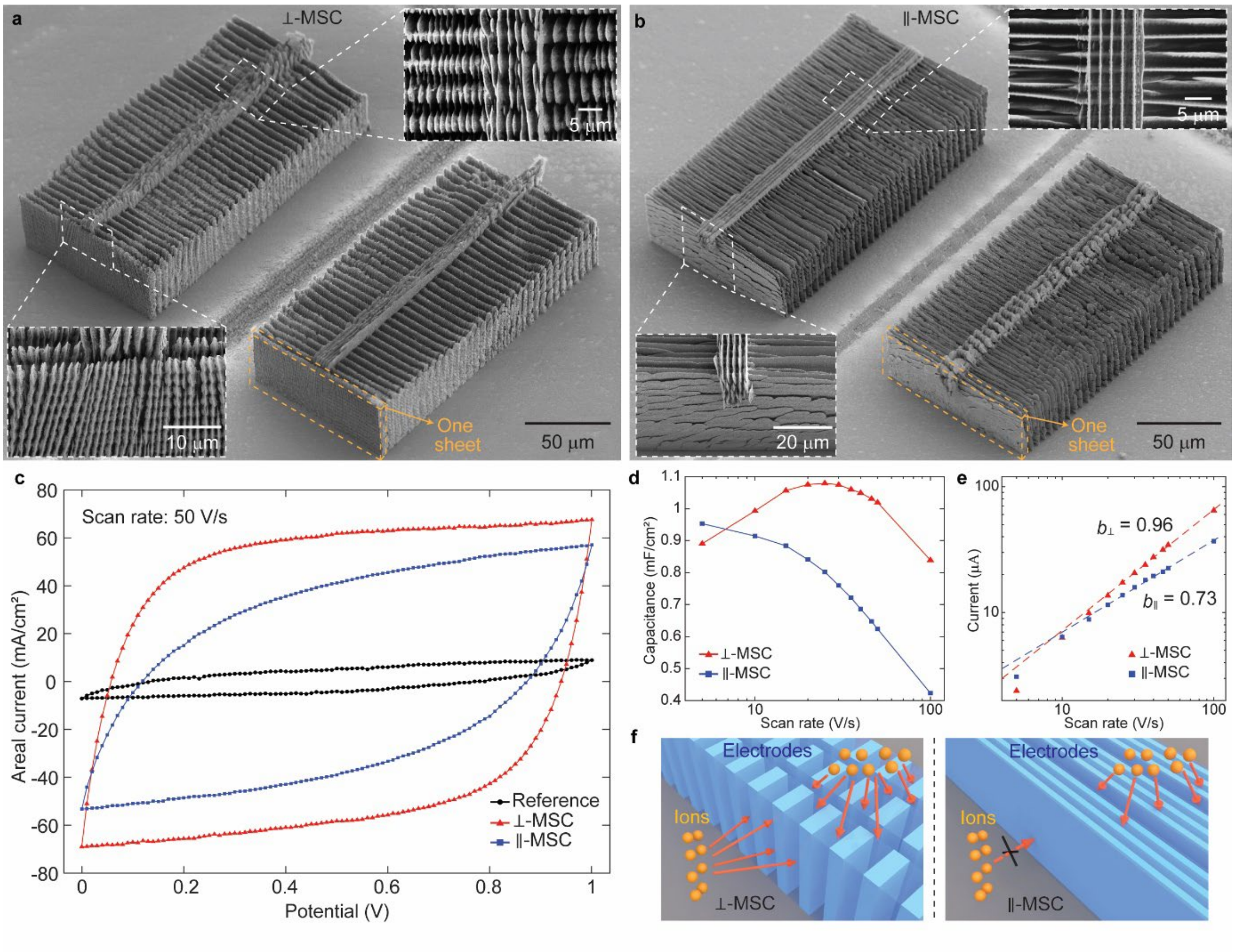

**Fig. 4. 3D-printed Si-rich glass micro-supercapacitors (MSCs) on silicon substrates. a, b,** SEM images of two different MSCs that were 3D-printed using identical laser writing paths but different laser polarizations, resulting in nanoscale structures perpendicular (⊥-MSC) and aligned (∥-MSC) to the extended direction of the vertical sheets to which they belonged, respectively. The insets are enlarged views of the nanoscale structures of the MSCs. **c,** Measured cyclic-voltammetry (CV) curves of the 3D-printed MSCs shown in **a** and **b,** and of a reference area without a 3D-printed structure at a scan rate of 50 V/s. **d, e,** Areal capacitances and currents, respectively, measured at different scan rates and extracted at the center of the potential window (0.5 V) of the corresponding CV curves. The voltage scanning range of the CV curves was fixed at 0 to 1 V. The dashed lines in **e** are exponential fits of the data for each MSC type. **f,** Schematic illustrations of ion transport in the cases of the ⊥-MSC and the ∥-MSC, respectively.

## Discussion

In this work, we presented an approach for 3D printing of hierarchical architectures made of inorganic Si-rich glass and composed of self-organized nanoscale structures, by femtosecond-laser direct writing inside hydrogen silsesquioxane (HSQ). The morphology of the nanoscale structures can be locally set in different parts of the 3D-printed architectures as desired by controlling the laser polarization and laser dose. Based on our morphological and material characterization of the 3D-printed structures, we propose that the underlying mechanism is the simultaneous occurrence of multiphoton crosslinking of HSQ and femtosecond-laser-induced nanograting formation inside glass materials. We demonstrated the utility of our approach by printing micro-supercapacitors with large surface areas, fast ion transport, and excellent cycling and temperature stability. Taken together, our observations pave new avenues for investigating the interactions of femtosecond-laser pulses and transparent materials and their applications in micro- and nanofabrication. Furthermore, 3D printing of inorganic hierarchical structures with tunable micro- and nano-morphology will enable innovative applications and research in important fields, including energy storage, nanophotonics, nanofluidics, nanoelectromechanical systems, and data storage. While beyond the scope of this work, multiple fundamental investigations of our 3D-printed glass structure are of interest for both research and applications. These include, for example, investigations of the detailed morphology of the nanogratings including the contact points and areas between the subordinate nanoplates, the structure of the glass at the atomic scale, and the internal elemental composition and distribution of the glass. Such investigations could provide insights contributing to a more comprehensive control over the 3D-printed structures and materials, a deeper understanding of the interplay of the two mechanisms involved in the 3D printing process, and the identification of other phenomena possibly involved such as the formation of silicon nanocrystals due to phase separations in HSQ[48]. Related insights could also provide crucial information for developing new glass-like resist materials that fulfill the conditions for the simultaneous occurrence of the two types of material modifications we observed in HSQ.

# Methods

## HSQ sample preparation

First, 20 wt% HSQ solution was prepared by dissolving HSQ powder (Applied Quantum Materials Inc.) in toluene (Honeywell Riedel-de-Haën™). Subsequently, the solution was repeatedly drop-casted onto the same substrate location to reach an HSQ layer with a thickness of at least 100 μm. Finally, the samples were left to dry at room temperature for at least eight hours before laser irradiation. For the 3D-printed micro-supercapacitors, silicon substrates (single-side-polished prime-grade silicon wafers (SIEGERT WAFER GmbH)) with a thermally grown 2 μm thick layer of silicon dioxide were used. Fused silica glass substrates (JGS2 optical-grade fused quartz (MicroChemicals)) were used for all other experiments. To demonstrate the compatibility of our approach with different substrate materials, two types of silicon substrates were used, including single-side-polished prime-grade silicon wafers (SIEGERT WAFER GmbH) and the same type of wafers with a thermally grown 2 μm thick layer of silicon-dioxide. We observed no differences in the 3D-printed Si-rich glass structures using these glass and silicon substrates.

## 3D direct writing by femtosecond-laser irradiation

A femtosecond laser source (Spirit 1040-4-SHG, Spectra-Physics of Newport Corporation) operating at a central wavelength of 1040 nm with a pulse duration of 298 fs and tunable repetition rate and pulse energy was used. A half-wave plate (10RP52-2, Newport Corporation) was installed to control the laser polarization. The laser was focused inside the HSQ or at the HSQ-substrate interface using an objective with a numerical aperture of 0.65 (Plan Achromat RMS40X, Olympus). The HSQ sample was fixed on a 3-axis linear motorized stage (XMS100, Newport) for moving the substrate with the HSQ relative to the spatially fixed laser focus to perform direct 3D laser writing. The laser pulse repetition rate, pulse energy, and laser scanning speed used in our experiments were 50 kHz, 125 nJ, and 100 μm/s, respectively, if not otherwise specified. The laser pulse energy was calculated by dividing the laser power measured by a thermopile detector (919P-010-16, Newport Corporation) by the laser pulse repetition rate. Two additional derived values are used in the analysis of the experimental results: (1) the temporal separation of laser pulses (in seconds) which is defined as the reciprocal of the laser pulse repetition rate and (2) the density of laser pulses (in pulses/μm) which is defined as the spatial pulse density in one laser-written line and calculated by dividing the laser pulse repetition rate by the laser scanning speed.

**Development of the 3D-printed structures**

The HSQ sample with the laser-irradiated 3D patterns was immersed in a developer of 0.1 M aqueous solution of potassium hydroxide (Sigma-Aldrich). Triton X-100 (LabChem Inc.) with 0.05 vol% was added to the developer as a surfactant to minimize the effects of bubbles formed in the development process. We left the samples in the developer until all the pristine HSQ was removed, which usually took at least 2 hours. Finally, the sample was removed from the developer, rinsed with isopropanol, and dried in air at room temperature.

**Material and morphology characterization of the 3D-printed glass**

The morphology of the developed samples was observed using scanning electron microscopy (SEM) (Ultra 55, Zeiss). The characterization of the dimensions of the self-forming nanogratings in HSQ was done by measuring the periodicity of the cracks and the thickness of the materials between two cracks observed in the cross-sectional SEM image of a femtosecond-laser-patterned HSQ sample that was mechanically cleaved. In total 25 periods (i.e., 26 cracks) were measured, and the periodicity and the thickness of the nanogratings are given in the format of the mean of the measured values ± one standard deviation of the measured values. Material characterization of the printed glass was carried out with confocal Raman and photoluminescence (PL) spectroscopy (alpha 300R, WITec) and energy-dispersive X-ray spectroscopy (EDS) (Aztec Ultim, Oxford Instruments). The Raman and photoluminescence spectroscope was equipped with a 405 nm wavelength laser. The laser power was manually set to 3 mW to maintain a sufficient signal intensity while preventing thermal damage to the samples. The collected light was guided through a single-mode fiber to a 300 mm ultrahigh-throughput spectrometer (UHTS 300, WITec). A 600 g/mm grating was used to disperse the collected light onto a CCD camera. The energy resolution of this system was at least 3 $cm^{-1}$, which is suitable for Raman and photoluminescence measurements. As for the EDS characterization, we prepared a sample containing pristine HSQ and 3D-printed glass on a fused-silica substrate. The sample was first coated with gold to prevent charging effects. Subsequently, for each measurement, an SEM image of the area of interest on the sample was taken and the sampling regions from which the EDS detector collected and integrated data were defined.

**Fabrication and characterization of micro-supercapacitors**

Si-rich glass electrode architectures with different nanoscale structures for micro-supercapacitor applications were designed and 3D-printed on silicon substrates with a thermally grown 2 μm thick silicon dioxide layer. Subsequently, the entire sample was coated with a titanium-nitride (TiN) layer with a thickness of 25 nm by atomic layer deposition (TFS200, BENEQ). This layer served as the electrode material and current collector for micro-supercapacitor applications. For each sample, the TiN between the two electrodes was removed by femtosecond laser ablation to isolate the electrodes electrically. Finally, the chosen electrolyte was drop-cast onto each sample and completely covered both electrodes. The cyclic volumetry (CV) characterization of the 3D-printed micro-supercapacitors was carried out in a two-electrode system using an electrochemical working station (Gamry Interface 1010E, Gamry Instruments Incorporation) with S-725-PRM micro-positioners (Signatone Corporation, Gilroy CA, USA). For measurements at room temperature, a standard probe station was used (S-1160, Signatone Corporation), and for high-temperature measurements, a hot-chuck probe station was used (S-1060R, Signatone Corporation). In the room-temperature measurements, the gel electrolyte of 1 M LiCl in poly(4-styrenesulfonic acid) (PSSH) aqueous solution was used, which was prepared by mixing 43 mg LiCl (99.5%, CAS number: 7447-41-8, VWR chemicals) in 1 mL PSSH solution (Mw~75,000, 18 wt% in $H_2O$, CAS number: 28210-41-5, Sigma-Aldrich). The high-temperature testing was performed in a temperature range from 50 °C to 250 °C with the ionic-liquid electrolyte of 1-butyl-3-methylimidazolium tetrafluoroborate (BMIM-BF4) (≥ 97.0% (HPLC), CAS number: 174501-65-6, Merck). For the CV curves, the current is presented in areal current, which is the measured current divided by the device footprint area. The 3D-printed micro-supercapacitors in this work had a footprint area of 56000 $\mu m^2$ including the electrodes and the gap between them. All capacitances provided in this work refer to areal capacitance $C_{\mathrm{areal}}$, which is calculated from the corresponding CV curves using the formula:

$$C_{\mathrm{areal}} = \frac{\int i dV}{2rAV_{\mathrm{window}}},$$

where $i$, $r$, $A$, $V_{\mathrm{window}}$ are the measured current, voltage scan rate, the footprint area of the device, and scanning potential window, respectively.

## Data availability

The data that support the findings of this work are available within this article and its Supplementary Information. Additional data related to this work may be requested from the corresponding author.

## Competing interests

A patent application (US 17/171,587) covering the methods in this work has been filed, with P.H., D.E.M., G.S., K.B.G., and F.N. as inventors and applicants. The remaining authors declare no competing interests.

## Acknowledgements

We thank Cecilia Aronsson for substrate dicing and assistance in the cleanroom. This work has been funded by Swedish Foundation for Strategic Research (SSF GMT14-0071 (G.S.) and SSF STP19-0014 (K.B.G.)), Swedish Research Council (Grant No. 2019-04731 (J.L.)), Wenner-Gren Foundation (UPD2020-0119 (D.E.M.)), Digitalization and Technology Research Center of the Bundeswehr under the project VITAL-SENSE (G.S.D. and O.H.) via German Recovery and Resilience Plan NextGenerationEU by the European Union, and European Union Horizon 2020 research and innovation program under the project Graphene Flagship Core 3 No. 881603 (G.S.D.), and SSLiP No. 101046693 (G.S.D.).

## Author contributions

P.H., D.E.M., K.B.G., and F.N. proposed the concept and conceived and designed the experiments to investigate the nanograting formation and develop the printing process. P.H. fabricated the structures and performed scanning electron microscopy and energy-dispersive X-ray characterizations. S.C., J.L., P.H., K.B.G., and F.N. designed, fabricated, and characterized the micro-supercapacitors, and analyzed the related data. O.H. and G.S.D. performed Raman and photoluminescence spectroscopy characterizations. All authors discussed the results and contributed to the manuscript writing.

# Supplementary Information
# 3D printing of hierarchical structures made of inorganic silicon-rich glass featuring self-forming nanogratings

Po-Han Huang[1], Shiqian Chen[2], Oliver Hartwig[3], David E. Marschner[1], Georg S. Duesberg[3], Göran Stemme[1], Jiantong Li[2], Kristinn B. Gylfason[1], and Frank Niklaus[1,*]

[1] Division of Micro and Nanosystems, School of Electrical Engineering and Computer Science, KTH Royal Institute of Technology, Stockholm 10044, Sweden
[2] Division of Electronics and Embedded Systems, School of Electrical Engineering and Computer Science, KTH Royal Institute of Technology, Kista 16440, Sweden
[3] Institute of Physics, EIT 2, Faculty of Electrical Engineering and Information Technology, University of the Bundeswehr Munich & SENS Research Center, Neubiberg 85577, Germany

[*] Corresponding author: Frank Niklaus (frank@kth.se)

**The Supplementary Information includes:**

**Supplementary Section 1. Material characterization**

To characterize the elementary composition and chemical bonds of the 3D-printed glass, we collected its energy-dispersive X-ray (EDS), Raman, and photoluminescence (PL) spectra. The EDS spectrum of the printed glass showed the peaks of silicon (Si), oxygen (O), and carbon (C), while EDS is incapable of detecting hydrogen by principle (Fig. 3a, b and Supplementary Fig. 3). Compared to the EDS spectrum of the pristine HSQ, the O/Si ratio and the C/Si ratio of the printed glass were both lower. The carbon content observed in the printed glass was at the same level as that observed in the fused silica substrate, and thus is considered as a common residual contamination originating from the environment. The Raman spectrum of the printed glass showed a dominant peak at ~496 $cm^{-1}$ which can originate from 4-membered Si-O rings in the silica glass network[1] and Si-Si bonds[2] (Fig. 3c). We also observed characteristic Raman features of fused silica glass such as the broad peak between 120 and 580 $cm^{-1}$, the 3-membered Si-O rings at ~605 $cm^{-1}$, and the Si-O-Si bending at ~805 $cm^{-1}$ in the spectrum of the printed glass[3]. Finally, we did not observe any feature around 2260 $cm^{-1}$ which is the spectral position of the signature Si-H peak of pristine cage-form HSQ (Supplementary Fig. 4). This indicates the occurrence of transformation of the pristine cage-form HSQ to the crosslinked network form upon femtosecond-laser exposure. In the PL spectrum of the printed glass, a broad and asymmetric peak between 1.3 and 2.6 eV was observed (Fig. 3d). The broad peak can be fitted by three peaks at ~1.6 eV from Si nanoclusters, at ~1.9 eV from non-bridging oxygen hole centers (NBOHC), and at ~2.55 eV possibly from hydrogen-related Si species[4–7]. EDS, Raman, and PL spectra of the glass structures printed with the laser polarization in parallel (Pol-X) and perpendicular (Pol-Y) to the writing direction showed no differences.

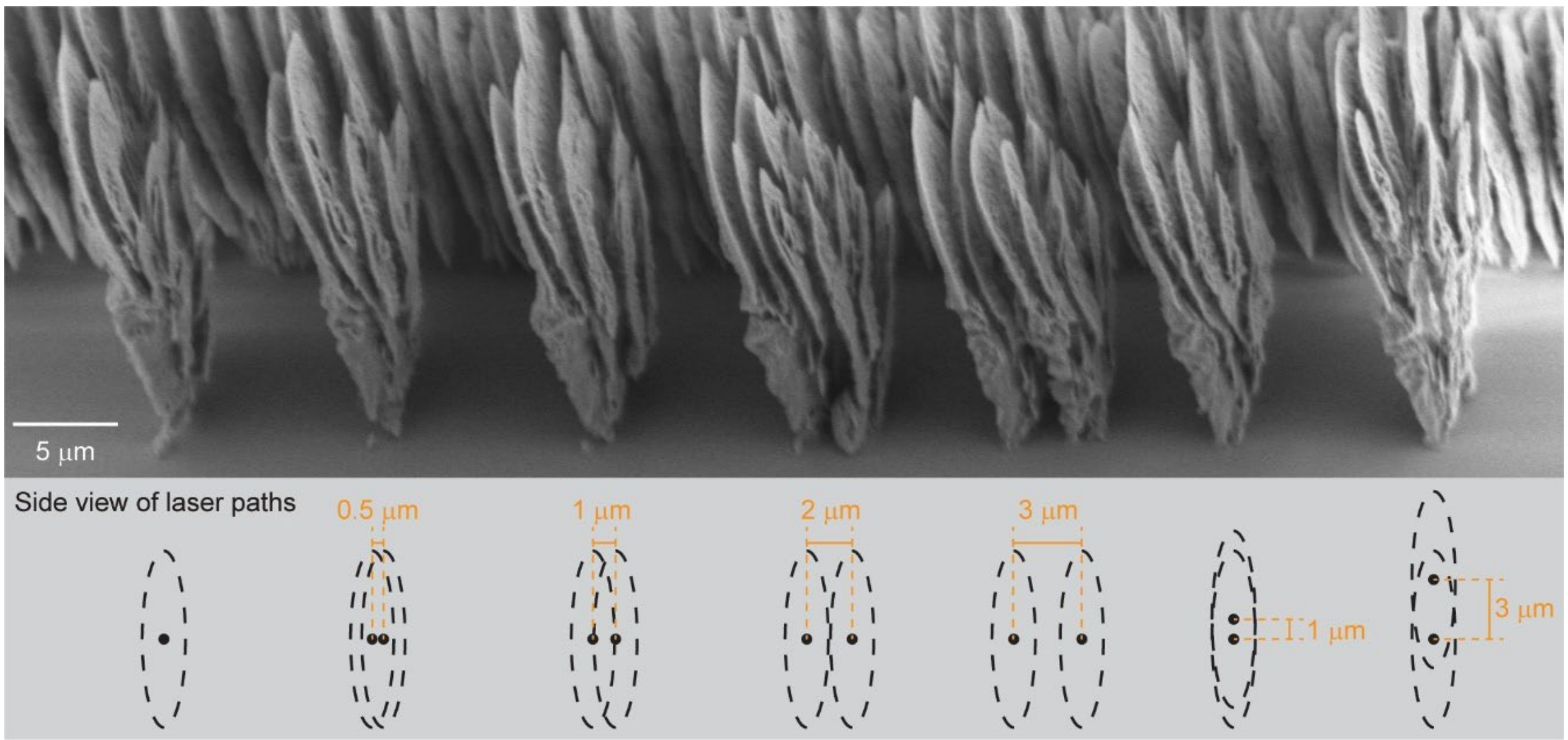

**Supplementary Fig. 1. Investigation of the contact regions between nanoplates in self-forming nanogratings.** Tilted-view SEM image of the side wall of seven independent 3D-printed line structures with schematics below showing the side view of their corresponding laser writing paths. For the structure to the left a single line laser writing path was used, while for each of the remaining structures two lines for the laser writing path were used. In the schematics, one line is represented by one dashed ellipse with one dot marking the center of the ellipse, and the distance between two lines in each two-line structure is annotated. We observed that the nanoplates in one nanograting (i.e., the nanograting formed within one laser written line) were connected at the bottom. In addition, when the distance between two lines of the laser writing path is larger than 2 μm, the contact region of the two self-forming nanogratings started to separate. For instance, two separate "tips" appeared in the structure comprising two lines laterally separated by 3 μm, and an elongated contact region appeared in the structure comprising two lines vertically separated by 3 μm. Further investigation of the evolution of the contact regions in nanogratings with increasing number of overlapping laser writing paths is required to fully clarify the sizes and positions of contact regions in large 3D-printed hierarchical structures featuring self-forming nanogratings.

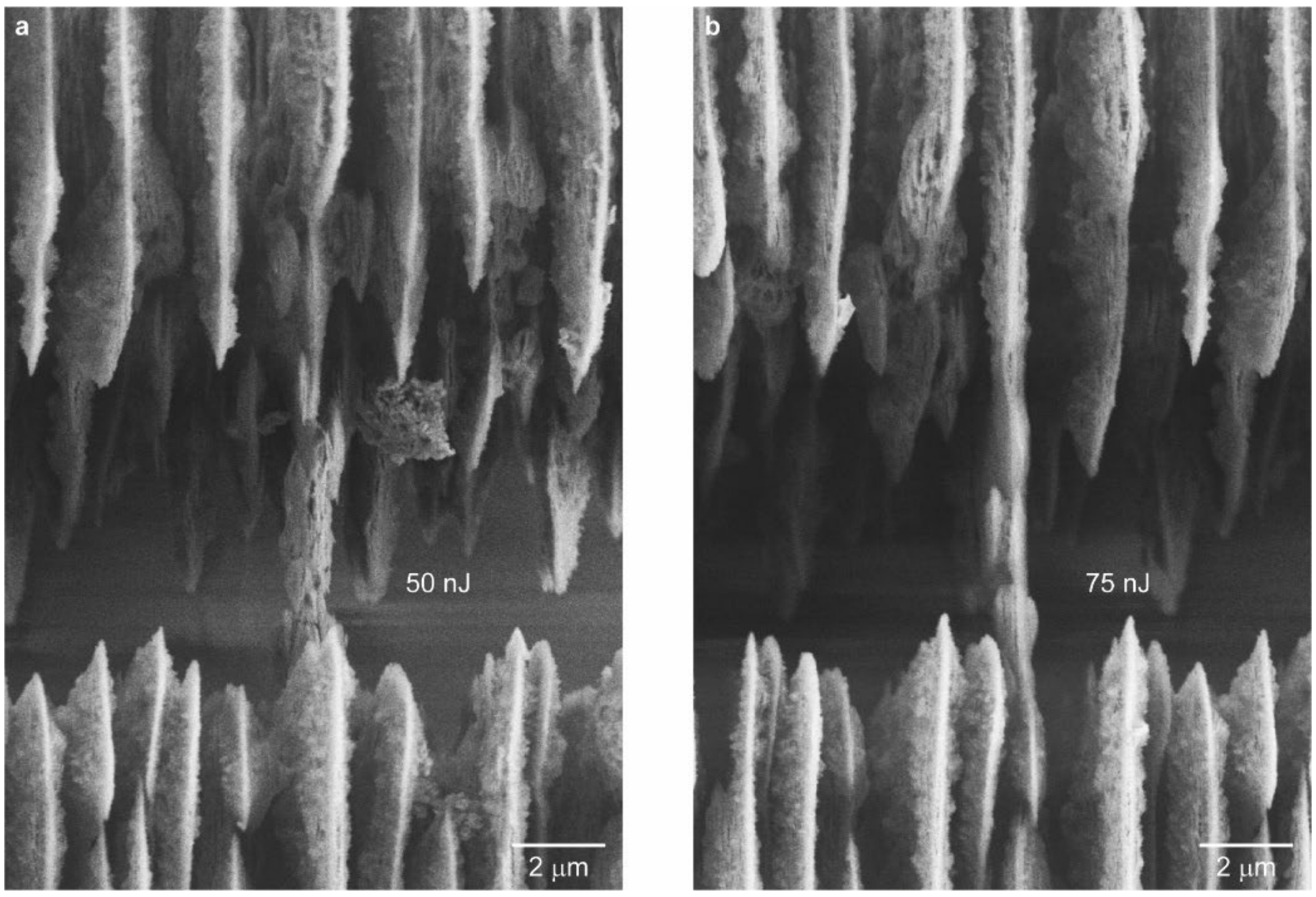

**Supplementary Fig. 2. Single-line structures with minimum width. a, b,** Enlarged SEM images of the 3D-printed suspended single line structures that were printed using laser pulse energies of 50 and 75 nJ, respectively, shown in Fig. 1g. Both structures were approximately 800 nm thick, while the structure in **a** was more porous than the structure in **b**, which indicates that the laser dose was not sufficient for the complete formation of the nanogratings in the former case.

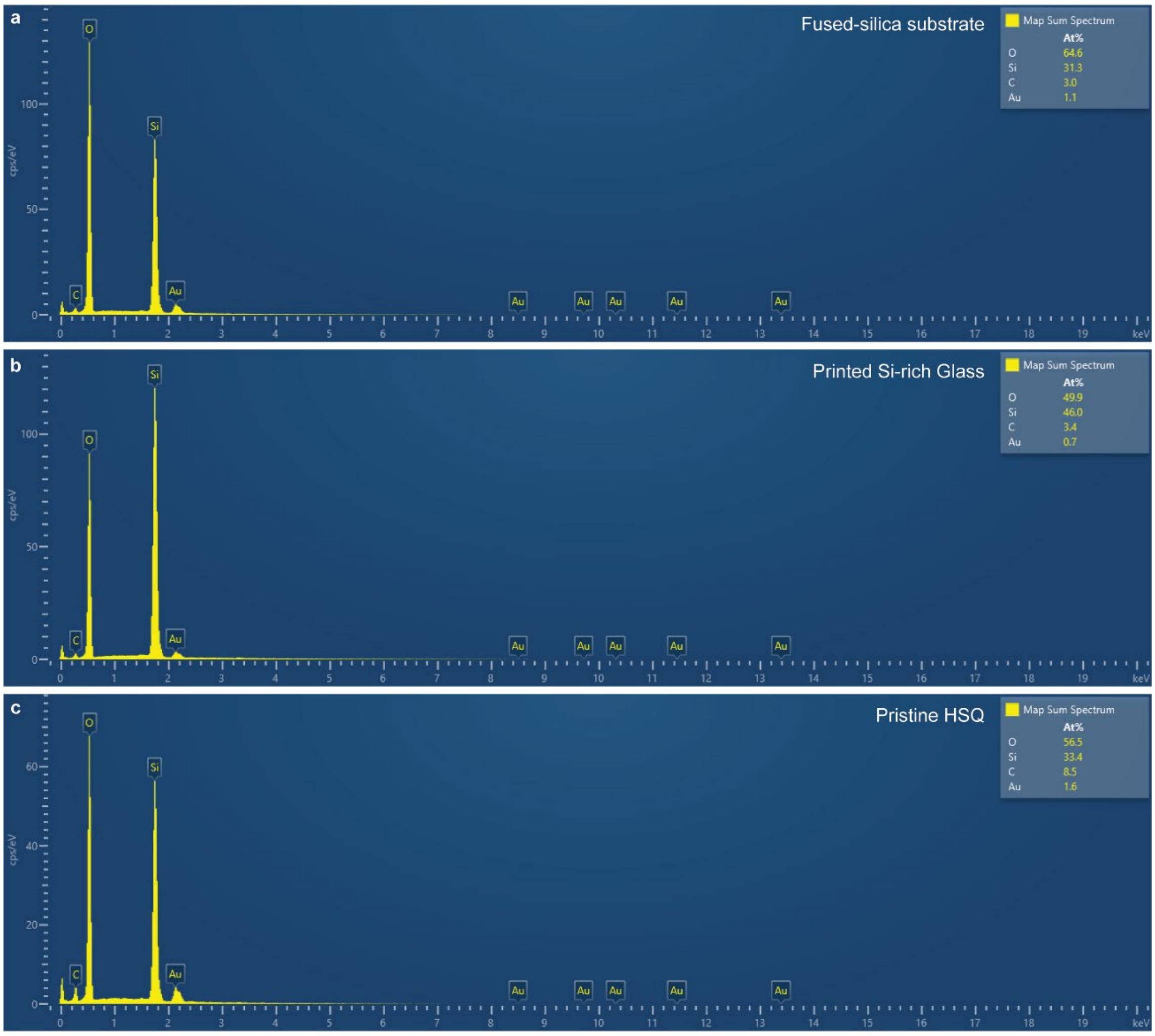

**Supplementary Fig. 3. Representative EDS spectra of fused-silica substrate, 3D-printed Si-rich glass, and HSQ, respectively.** The spectra were measured of the fused-silica substrate (**a**), 3D-printed Si-rich glass (**b**), and pristine HSQ (**c**) on the same sample. The gold peaks observed in the spectra originated from the gold coating on the sample to prevent charging effects during the measurements.

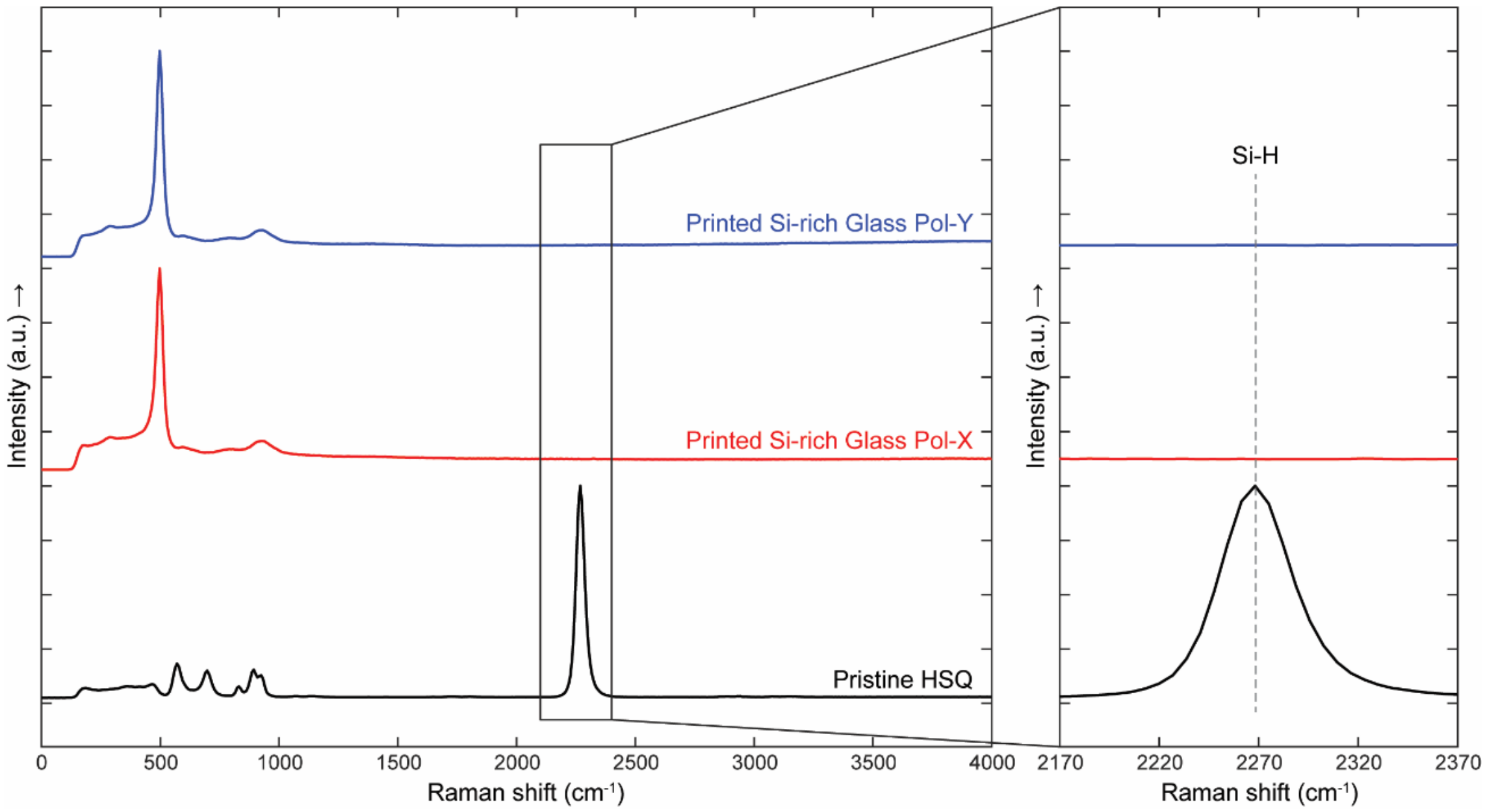

**Supplementary Fig. 4. Extended Raman spectra of the 3D-printed Si-rich glass and the pristine HSQ.** The spectra were measured of a pristine HSQ sample (black) and two Si-rich glass samples printed with the laser polarization in parallel (Pol-X, red) and perpendicular (Pol-Y, blue) to the laser writing direction, respectively. An enlarged view of the spectra around the Raman shifts of 2260 $cm^{-1}$, which is the spectral position of the signature Si-H peak of pristine HSQ, is provided to show that the Si-H Raman feature was not present in the 3D-printed Si-rich glass.

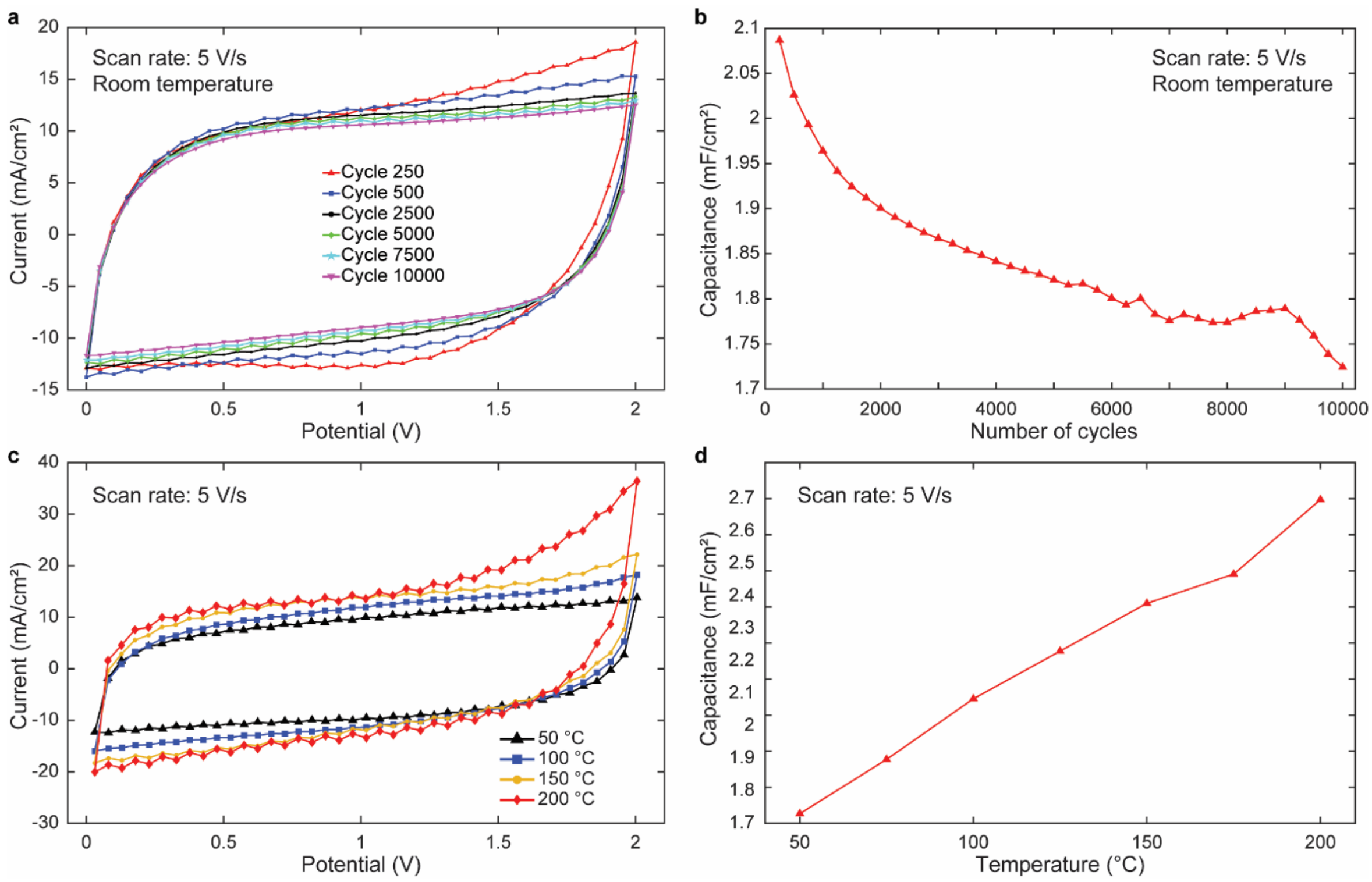

**Supplementary Fig. 5. Long-cycling and high-temperature experiments of a 3D-printed Si-rich glass micro-supercapacitor with an ionic-liquid electrolyte. a,** Measured cyclic-voltammetry (CV) curves of the 3D-printed micro-supercapacitor after different charging-discharging cycles at room temperature. **b,** Areal capacitances extracted from the corresponding measured CV curves of the 3D-printed micro-supercapacitor after different charging-discharging cycles at room temperature. **c,** Measured CV curves of the 3D-printed micro-supercapacitor at different temperatures. **d,** Areal capacitances extracted from the corresponding measured CV curves of the 3D-printed micro-supercapacitor at different temperatures.

**Supplementary Table 1. Comparison of high-scan-rate supercapacitors in literature.**

| Supercapacitor | Capacitance | Voltage window | Electrolyte | Reference |
|---|---|---|---|---|
| MXene (electrode only) | 100 mF $cm^{-2}$ @50 V $s^{-1}$ | 0-0.6 V | Liquid $H_2SO_4$ | 8 |
| r-GO/TiO | 0.35 mF $cm^{-2}$ @50 V $s^{-1}$ | 0-0.8 V | PVA/$H_2SO_4$ | 9 |
| Graphene | 0.021 mF $cm^{-2}$ @50 V $s^{-1}$ | 0-1 V | PVA/$H_2SO_4$ | 10 |
| Graphene | 0.23 mF $cm^{-2}$ @10 V $s^{-1}$<br>0.15 mF $cm^{-2}$ @100 V $s^{-1}$ | 0-1 V | PVA/$H_2SO_4$ | 11 |
| MXene/PEDOT | 0.62 mF $cm^{-2}$ @50 V $s^{-1}$ | 0-0.8 V | c-PVA/$H_2SO_4$ | 12 |
| Graphene/Au | 0.11 mF $cm^{-2}$ @50 V $s^{-1}$ | 0-0.8 V | PVA/$H_2SO_4$ | 13 |
| Graphene | 0.86 mF $cm^{-2}$ @50 V $s^{-1}$ | 0-1 V | PVA/$H_2SO_4$ | 14 |
| Graphene | 0.082 mF $cm^{-2}$ @10 V $s^{-1}$<br>0.032 mF $cm^{-2}$ @100 V $s^{-1}$ | 0-1 V | PVA/$H_2SO_4$ | 15 |
| 2D rGO/Au<br>3D rGO/Au | 1 mF $cm^{-2}$ @50 V $s^{-1}$<br>3 mF $cm^{-2}$ @50 V $s^{-1}$ | 0-1 V | PVA/$H_2SO_4$ | 16 |
| N/B co-doped graphene | 0.063 mF $cm^{-2}$ @50 V $s^{-1}$ | 0-1 V | PVA/$H_2SO_4$ | 17 |
| PANI/Graphene | 0.34 mF $cm^{-2}$ @50 V $s^{-1}$ | 0-0.8 V | Liquid $H_2SO_4$ | 18 |
| Graphene | 0.021 mF $cm^{-2}$ @50 V $s^{-1}$ | 0-1 V | PVA/$H_2SO_4$ | 19 |
| S doped Graphene | 0.06 mF $cm^{-2}$ @40 V $s^{-1}$ | 0-1 V | PVA/$H_2SO_4$ | 20 |
| Graphene Nanoribbons | 1.64 mF $cm^{-2}$ @10 V $s^{-1}$<br>0.7 mF $cm^{-2}$ @100 V $s^{-1}$ | 0-1 V | PVA/$H_2SO_4$ | 21 |
| MXene | 1.96 mF $cm^{-2}$ @20 V $s^{-1}$<br>1.6 mF $cm^{-2}$ @80 V $s^{-1}$ | 0-0.6 V | PVA/$H_2SO_4$ | 22 |
| Carbon onions | 0.9 mF $cm^{-2}$ @100 V $s^{-1}$ | 0-3 V | Et4NBF4/anhydrous propylene carbonate electrolyte | 23 |
| **TiN/Si-rich glass nanoplates** | **1 mF $cm^{-2}$ @50 V $s^{-1}$**<br>**2.09 mF $cm^{-2}$ @5 V $s^{-1}$** | **0-1 V**<br>**0-2 V** | **PSSH/LiCl**<br>**BMIM-BF4** | **This work** |